\documentclass[aps,prd,twocolumn,showpacs,nofootinbib]{revtex4-1}
\usepackage[utf8]{inputenc}
\usepackage{latexsym}
\usepackage{amsmath,amsfonts}
\usepackage{amsbsy}
\usepackage{mathrsfs}
\usepackage{color}
\usepackage{psfrag}
\usepackage{enumerate}
\usepackage{amsmath,amssymb,calc,amsfonts}
\usepackage{latexsym}
\usepackage[colorlinks=true,linkcolor=blue,urlcolor=blue,citecolor=red]{hyperref}
\usepackage{amsmath}
\usepackage{amsfonts}
\usepackage{amssymb}
\usepackage{appendix}
\usepackage{latexsym}
\usepackage{amsfonts}
\usepackage{t1enc}
\usepackage{times}
\usepackage{xmpmulti}
\usepackage{color}
\usepackage{bm}
\usepackage{hyperref}

\font\tenscr=rsfs10 scaled1100
\font\sevenscr=rsfs7 
\font\fivescr=rsfs5 
\skewchar\tenscr='177
\skewchar\sevenscr='177
\skewchar\fivescr='177
\newfam\scrfam
\textfont\scrfam=\tenscr
\scriptfont\scrfam=\sevenscr
\scriptscriptfont\scrfam=\fivescr

\def\scri{{\fam\scrfam I}}
\def\scrm{{\fam\scrfam M}}

\begin{document}

\title{Asymptotic structure of spacetime and the Newman-Penrose formalism: a brief review  }

\author{
\small L. A. G\'omez L\'opez$^\dagger$ and \small G. D. Quiroga$^\dagger$\\
\em \normalsize \em \small ${}^\dagger$ GIRG, Escuela de F\'isica, Universidad Industrial de Santander \\
\normalsize \em \small A. A. 678, Bucaramanga, Colombia\\
\normalsize \em \small gquiroga@uis.edu.co
}

\begin{abstract}
A brief review about the Newman-Penrose formalism and the asymptotic structure of the spacetime is given. The goal of this review is to describe the latest developments in these topics and make a summary of the most important articles published by Newman and collaborators. Additionally, we discuss some aspects of this approach, and we compute the spin coefficients and the Weyl scalars for a general stationary axisymmetric spacetimes in a tetrad basis different from that defined by the principal null geodesic directions.
\\

\noindent \emph{Keywords:} General Relativity. Asymptotically flat spacetime.
\\

\noindent PACS numbers: 04.20.-q, 04.20.Ha
\end{abstract}

\maketitle

\section{Introduction}\label{intro}
The asymptotic structure of spacetime and the Newman-Penrose (NP) formalism \cite{np} are powerful tools for analyzing the behavior of isolated systems and compact sources in General Relativity (GR). This formalism has found many applications in the last past years in different areas of the GR, and also in numerical relativity. From the mathematical point of view, the NP approach is a special case of the tetrad formalism, where a set of four null vectors is chosen to form a basis at each point of the spacetime. Then, several complex scalars are introduced to describe the dynamic of the gravitational fields at infinity, and the evolution equations of these complex scalars, i.e. the Bianchi identities, are written. Furthermore, these vectors are appropriate for studying the properties of congruences of null geodesics \cite{ANK}, such congruences naturally arise in the context of the propagation of the gravitational and the electromagnetic radiation \cite{quiroga2017sources}.

As has been mentioned before, the NP formalism has been used in several aspects of analytical and computational relativity: recently, in collaboration with Kozameh, this formalism was used to describe the behavior of compact sources, such as Black Holes collisions and close binary coalescences, linking the dynamics of the system to the gravitational radiation emitted. Moreover, we find some similarities and several differences with other approaches like the Post-Newtonian equations \cite{KQ2}. Furthermore, the NP approach is used for extracting gravitational waves from numerical simulation via the computation of one of the five Weyl complex scalars. Although the NP formalism is primarily focused on studying the properties of radiation of isolated systems, this formulation is a very useful framework for constructing and investigating exact solutions of the four-dimensional General Relativity. Additionally, there are attempts to generalize this approach to higher dimensions \cite{pravdova2008newman}. Particularly, the method is very powerful when the spacetime is algebraically special according the Petrov classification \cite{hall2004symmetries}, since some complex scalars can be set to zero by choosing an appropriate tetrad aligned with the outgoing tangent vector fields of null radial geodesics.

In this article, we propose to make a short review about the NP approach, giving some applications of this approach and explaining the physical interpretation of the most important scalars introduced by Newman and Penrose. This article is organized as follows: In section \ref{sec.2} we give the geometric notion of an asymptotically flat spacetime, and we introduce a set of null vectors to define the complex Weyl scalars, the Maxwell scalars, and the twelve spin coefficients used in the NP formalism. In the section \ref{sec1.5}, we reduce all the equations and definitions to a particular set of coordinates, usually called Bondi coordinates, and we introduce the notion of the Mass and the Bondi linear momentum. The BMS group and the transformation between different families of null cuts is shown in section \ref{sec.4}, while the physical meaning of the Weyl scalars is discussed in sections \ref{sec.5}. In section \ref{sec:2}, we use the NP approach to analyze a more general stationary axisymmetric spacetime, where we compute all the NP quantities, and we apply the resulting equations to some familiar spacetimes. Finally, we conclude this work giving some finals remarks and comments.

\section{Asymptotically flat spacetime} \label{sec.2}
The notion of asymptotically flat spacetime is an adequate tool to analyze the gravitational and electromagnetic radiation coming from an arbitrary compact source.
During the 60s, Bondi, Sachs and collaborators \cite{BBM,sachs,sachs3} used a system of canonical coordinates to define the mass, momentum and gravitational radiation. Then, Penrose defines the notion of asymptotically flat spaces (or asymptotically simple), using the idea of re-scale the metric by an appropriate factor, usually called conformal factor, which is appropriately chosen to decay to zero at infinity \cite{Pen}. The geometric notion of Penrose can be summarized in the following definition
\bigskip

\textbf{\emph{Definition:}}
A spacetime $(\scrm,g_{ab})$  is called asymptotically simple, if the curvature tensor goes to zero as we approach to infinity in the future direction of the null geodesics of the spacetime. These geodesics end up in what is called the future null infinity $\scri^+$. A future null asymptote is a manifold $\hat \scrm$ with boundary $\scri^+ \equiv \partial \hat\scrm$ together with a smooth Lorentzian metric $\hat{g}_{ab}$, and a smooth function $\Omega$ on $\hat\scrm$ satisfying the following
\begin{itemize}
\renewcommand\labelitemi{$\circ$}
  \item $\hat{\scrm}=\scrm \cup \scri^+$
  \item On $\scrm$, $\hat{g}_{ab}=\Omega ^2 g_{ab}$ with $\Omega >0$
  \item At $\scri^+$, $\Omega=0$, $n_a \equiv \partial _a \Omega \neq 0$ and $\hat {g}^{ab}n_a n_b =0$
\end{itemize}
From this geometric definition, it is possible to show that,
\begin{enumerate}
  \item If $\scrm$ satisfy the vacuum Einstein equations near $\scri$ then $\scri$ it is a null boundary. Also $\scri$ consists of two disjoint parts $\scri^+$ and $\scri^-$, each topologically $S^2 \times R$.
  \item The Weyl tensor ${C_{abc}}^d$ vanishes at $\scri$ and the peeling assumption \cite{sachs} establishes the way it approaches to zero. Furthermore, the Weyl tensor is conformally invariant, i.e  the tensor $\hat{C}_{abc}^{\ \ \ d}$ constructed using the re-scaled metric $\hat{g}_{ab}$ and ${C_{abc}}^d$ are the same at null infinity \cite{EspWitt}.
\end{enumerate}
We begin the study of the null infinity properties, introducing a coordinate system in the neighborhood of $\scri^+$, which we will label as $(u,r,\zeta,\bar\zeta)$ \cite{nu}. In this system, the time $u$ represents a family of null surfaces, $r$ is the affine parameter along the null geodesics of the constant $u$ surfaces, and $\zeta=e^{i \phi} cot \frac{\theta}{2}$, the complex stereographic angle labeling the null geodesics of $\scri^+$. We reach $\scri^+$ taking $r \rightarrow \infty $, thus the null infinity is described by the remaining coordinates $(u, \zeta,\bar\zeta)$. Now, the two-surface metric becomes,
\begin{equation}\label{metricasphy}
ds^2= -\frac{4r^2 d\zeta d\bar\zeta}{P^2},
\end{equation}
making the usual choice of $\Omega=r^{-1}$ as the conformal factor, eq. (\ref{metricasphy}) gives the induced metric on $\scri^+$,
\begin{equation} \label{metricascri}
d\hat{s}^2=-\frac{4d\zeta d\bar \zeta}{P^2}.
\end{equation}
Here $P(u,\zeta,\bar\zeta)$ is a strictly positive function which depends on the framework choice.

\subsection{The Newman-Penose formalism} \label{sec1.1}
The Newman-Penrose (NP) formalism \cite{np,Pen} is a tetrad formalism based on a set of four null vectors. Associated with the coordinates  $(u,r,\zeta,\bar\zeta)$ one can introduce a tetrad of two real vectors denoted by $l^a$ and $n^a$, and two complex conjugate vectors $m^a$ and $\bar{m}^a$, where the first co-vector is defined by \cite{ntod}
\begin{equation}
l_{a}=\nabla _{a}u,
\end{equation}
so $l^{a}=g^{ab}\nabla_a u$ is a null vector tangent to the generators of $u=const.$ The remaining vectors are required to satisfy the following orthogonality conditions
\begin{align}
l^a n_a&=-m^a\bar{m}_a=1,\\
l^am_a=l^a\bar{m}_a&=n^am_a=n^a\bar{m}_a=0,\\
l^al_a=n^an_a&=m^am_a=\bar{m}^a\bar{m}_a=0.
\end{align}
The indices can be raised and lowered using the global metric $g_{ab}$ which, in terms of null vectors, can be written as
\begin{equation} \label{metrica}
g_{ab}=l_{a}n_{b}+n_{a}l_{b}-m_{a}\bar{m}_{b}-\bar{m}_{a}m_{b}.
\end{equation}
Also, the metric (\ref{metrica}) can be expressed in condensed notation as
\begin{equation}
g^{ab}=\eta^{\mu \nu}{\lambda^{a}}_{\mu}{\lambda^{b}}_{\nu},
\end{equation}
where $\mu$ is the tetrad index, which are raised and lowered using the flat metric $\eta^{\mu \nu}$ and $\eta_{\mu \nu}$, and where ${\lambda^a}_{\mu}$ is given by
\begin{equation}
{\lambda^a}_{\mu}=(l^{a}, n^{a}, m^{a}, \bar{m}^{a}), \qquad \mu=1,2,3,4
\end{equation}
and
\begin{equation}
\eta ^{\mu \nu}=\left(
\begin{array}{cccc}
0 & 1 & 0 & 0 \\
1 & 0 & 0 & 0 \\
0 & 0 & 0 & -1 \\
0 & 0 & -1 & 0%
\end{array}%
\right),
\end{equation}
It is important to note that the tetrad has certain freedoms such as
\begin{enumerate}[(a)]
  \item To perform null rotations about $l_a$:
\end{enumerate}
\begin{eqnarray}
l^{a} &\rightarrow& l^{a}, \\
m^{a} &\rightarrow& m^{a}+B l^{a}, \\
n^{a} &\rightarrow& n^{a}+\bar{B} m^{a}+B \bar{m}^{a}+B\bar{B}l^{a}.
\end{eqnarray}
\begin{enumerate}[(b)]
\item A null rotations about $n_a$:
\end{enumerate}
\begin{eqnarray}
l^{a} &\rightarrow& l^{a}+\bar{B} m^{a}+B \bar{m}^{a}+B\bar{B}n^{a},  \\
m^{a} &\rightarrow& m^{a}+B n^{a}, \\
n^{a} &\rightarrow& n^{a}.
\end{eqnarray}
\begin{enumerate}[(c)]
\item To rotate $m^a$ while keeping $l^a$ and $n^a$ fixed:
\end{enumerate}
\begin{equation} \label{rot}
m^{a} \rightarrow e^{i\lambda}m^{a} \qquad for \ a \ real \ \lambda.
\end{equation}
This freedom introduces the notion of spin weight, a quantity $\eta$ that under (\ref{rot}) transforms as,
\begin{equation}
\eta \rightarrow e^{is\lambda}\eta,
\end{equation}
is said to have spin weight $s$ \cite{ntod}. One can also define the spin weighted differential operators $\eth$ and $\bar\eth$ as follows,
\begin{eqnarray}
\eth f&=&P^{1-s}\frac{\partial (P^{s}f)}{\partial \zeta }, \label{ethG}\\
\bar{\eth }f&=&P^{1+s}\frac{\partial (P^{-s}f)}{\partial \bar{\zeta}}, \label{ethbG}
\end{eqnarray}
where $P$ is the factor that involves the metric of spacetime (\ref{metricascri}). The operators $\eth$ and $\bar\eth$ raise and lower the spin weight by one respectively.
\begin{enumerate}[(d)]
\item Finally, since the choice of the coordinate system is not unique, so we can make a different choice from the original $u=const.$ cut, i.e,
\end{enumerate}
\begin{equation}
u \rightarrow \hat{u}=T(u,\zeta,\bar\zeta).
\end{equation}
Under this choice, the tetrad system transforms as
\begin{eqnarray}
l^{a} &\rightarrow& \dot T \left( l_{a}+\frac{\eth T}{\dot T r}\bar{m}_{a}+\frac{\bar{\eth}T}{\dot T r}m_{a}+\frac{\eth T \bar{\eth}T}{\dot T^2 r^2}n_{a}\right), \label{la}\\
m^{a} &\rightarrow& m^{a}+\frac{\eth T}{\dot T r} n^{a}, \\
n^{a} &\rightarrow& \dot T^{-1} n^{a}. \label{na}
\end{eqnarray}
These equations give the expansion between two tetrads, and will be of great importance in the following sections.

\subsection{Spin coefficient formalism}\label{SC}
In the Newman-Penrose formalism \cite{np}, one introduces twelve complex spin coefficients (SC), five complex functions encoding the Weyl tensor, three complex Maxwell scalars, and ten functions encoding the Ricci tensor in the tetrad basis. These complex functions are the primary quantities used in this asymptotic formulation of the General Relativity. It is well known that this formulation is suitable to describe isolated systems and analyze the gravitational radiation emitted by compact sources \cite{KQ2}.
In this article, we will focus on the general form of the asymptotically flat solutions of the Einstein-Maxwell equations. Particularly, in this section, we will introduce all these quantities, starting with the Ricci rotation coefficients $\gamma_{\mu \nu \rho}$ defined by,
\begin{equation}
\gamma_{\mu \nu \rho}={\lambda ^{a}}_{\rho}{\lambda ^{b}}_{\nu}\nabla _{a}\lambda_{b \mu},
\end{equation}
where $\mu,\nu,\rho=1,2,3,4$ are tetrad indexes, and where the Ricci rotation coefficients satisfy,
\begin{equation}
\gamma _{\mu \nu \rho}=-\gamma _{\nu \mu \rho}.
\end{equation}%
Then, the spin coefficients (SC) are defined as combinations of the $\gamma_{\mu \nu \rho}$ by the following equations,
\begin{eqnarray}
\alpha  &=&\frac{1}{2}(\gamma _{124}-\gamma _{344}),\quad \lambda =-\gamma
_{244},\quad \kappa =\gamma _{131},  \nonumber  \label{spincoef} \\
\beta  &=&\frac{1}{2}(\gamma _{123}-\gamma _{343}),\quad \mu =-\gamma
_{243},\quad \rho =\gamma _{134}, \\
\gamma  &=&\frac{1}{2}(\gamma _{122}-\gamma _{342}),\quad \nu =-\gamma
_{242},\quad \sigma =\gamma _{133},  \nonumber \\
\varepsilon  &=&\frac{1}{2}(\gamma _{121}-\gamma _{341}),\quad \pi =-\gamma
_{241},\quad \tau =\gamma _{132}.  \nonumber
\end{eqnarray}
Since the spacetime is assumed to be empty in a neighborhood of the null infinity, the gravitational field is given by the Weyl tensor. Using the available tetrad, one defines five complex scalars in the following way
\begin{eqnarray}
\psi _{0}=-C_{abcd}l^{a}m^{b}l^{c}m^{d}; \quad \psi _{1}=-C_{abcd}l^{a}n^{b}l^{c}m^{d}, \nonumber\\
\psi _{2}=-\frac{1}{2}(C_{abcd}l^{a}n^{b}l^{c}n^{d}-C_{abcd}l^{a}n^{b}m^{c}{\bar{m}}^{d}), \\
\psi _{3}=C_{abcd}l^{a}n^{b}n^{c}{\bar{m}}^{d}; \quad \psi _{4}=-C_{abcd}n^{a}{\bar{m}}^{b}n^{c}{\bar{m}}^{d}. \nonumber
\end{eqnarray}
When an electromagnetic field is present we can introduce the Maxwell tensor $F_{ab}=\partial_a A_b-\partial_b A_a$,  from where we compute three complex Maxwell scalars given by
\begin{eqnarray}
\phi _{0}&=&F_{ab}l^{a}m^{b}, \quad \phi _{1}=\frac{1}{2}F_{ab}(l^{a}n^{b}+m^{a}{\bar{m}}^{b}), \nonumber \\
\phi _{2}&=&F_{ab}n^{a}{\bar{m}}^{b}.
\end{eqnarray}
From the ``peeling'' assumption introduced by Sachs \cite{sachs}, one can obtain the asymptotic behavior of the Weyl and Maxwell scalars, and the spin coefficients for any asymptotically flat spacetime. These scalars fall to zero as inverse powers of $r$ in the following way \cite{ntod}
\begin{align}
\psi _{0}&=\psi _{0}^{0}r^{-5}+O(r^{-6}), \quad \psi _{1}=\psi _{1}^{0}r^{-4}+O(r^{-5}), \nonumber \\
\psi _{2}&=\psi _{2}^{0}r^{-3}+O(r^{-4}), \quad \psi _{3}=\psi _{3}^{0}r^{-2}+O(r^{-3}), \nonumber \\
\psi _{4}&=\psi _{4}^{0}r^{-1}+O(r^{-2}), \quad \phi _{0}=\phi _{0}^{0}r^{-3}+O(r^{-4}),\\
\phi _{1}&=\phi _{1}^{0}r^{-2}+O(r^{-3}), \quad \phi _{2}=\phi _{2}^{0}r^{-1}+O(r^{-2}). \nonumber
\end{align}
Also, the null vectors in the NP formalism can be written as
\begin{eqnarray}
l &=&l^{a}\frac{\partial }{\partial x^{a}}=\frac{\partial }{\partial r}, \nonumber\\
n &=&n^{a}\frac{\partial }{\partial x^{a}}=\frac{\partial }{\partial u}+U%
\frac{\partial }{\partial r}+X^{\zeta }\frac{\partial }{\partial \zeta }+X^{ %
\bar{\zeta}}\frac{\partial }{\partial \bar{\zeta}}, \nonumber\\
m&=&m^{a}\frac{\partial }{\partial x^{a}}=\omega \frac{\partial }{\partial r
}+\xi ^{\zeta }\frac{\partial }{\partial \zeta }+\xi ^{ \bar{\zeta}}\frac{%
\partial }{\partial \bar{\zeta}}, \\
\bar{m}&=&\bar{m}^{ a}\frac{\partial }{\partial x^{a}}=\bar{\omega}\frac{%
\partial }{\partial r}+\bar{\xi}^{ \zeta }\frac{\partial }{\partial \zeta }+%
\bar{\xi}^{ \bar{\zeta}}\frac{\partial }{\partial \bar{\zeta}}, \nonumber
\end{eqnarray}
where
\begin{eqnarray*}
\xi ^{k} &=&\xi ^{0k}r^{ -1}-\sigma ^{0}\bar{\xi}^{ 0k}r^{ -2}+O(r^{ -3}) \quad with \quad k=\zeta,\bar\zeta \nonumber\\
\omega  &=&\omega ^{ 0 }r^{ -1}-\left( \sigma ^{ 0}\bar{\omega}^{ 0}+\frac{\psi
_{1}^{0}}{2}\right) r^{-2}+O(r^{-3}), \\
X^{ k} &=&(\psi _{1}^{0}\bar{\xi}^{ 0k}+\bar{\psi}_{1}^{0}\xi^{0k})(6r^3)^{-1}+O(r^{-4}), \nonumber\\
U &=&U^{0}-(\gamma ^{ 0}+\bar{\gamma}^{0})r-(\psi _{2}^{0}+\bar{\psi}%
_{2}^{0})(2r)^{-1}+O(r^{ -2}). \nonumber
\end{eqnarray*}
Now, the spin coefficients are given by \cite{nu},
\begin{eqnarray} \label{spincoef-des-r}
\kappa &=&\pi =\varepsilon =0, \qquad \rho=\bar{\rho }, \qquad \tau=\bar{\alpha }+\beta, \nonumber\\
\rho &=&-r^{ -1}-\sigma ^{0}\bar{\sigma }^{0}r^{ -3}+O(r^{ -5}), \nonumber\\
\sigma &=&\sigma ^{ 0}r^{ -2}+[(\sigma ^{ 0})^{2}\bar{\sigma }^{ 0}-\psi
_{0}^{0}/2]r^{ -4}+O(r^{ -5}), \nonumber\\
\alpha &=&\alpha ^{ 0}r^{ -1}+O(r^{ -2}), \nonumber\\
\beta&=&\beta ^{ 0}r^{ -1}+O(r^{ -2}), \\
\gamma &=&\gamma ^{ 0}-\psi _{2}^{0 }(2r^{ 2})^{-1}+O(r^{ -3}), \nonumber\\
\mu &=&\mu ^{ 0}r^{\ -1}+O(r^{ -2}), \nonumber\\
\lambda &=&\lambda ^{ 0}r^{ -1}+O(r^{ -2}), \nonumber\\
\nu&=&\nu ^{ 0}+O(r^{ -1}), \nonumber
\end{eqnarray}
where,
\begin{eqnarray}
\xi ^{0\zeta }&=&-P, \qquad \bar{\xi }^{ 0\zeta }=0, \nonumber\\
\xi ^{ 0\bar{\zeta }}&=&0, \qquad \bar{\xi }^{ 0\bar{\zeta }}=-P ,\nonumber\\
\alpha ^{ 0}&=&-\bar{\beta }^{ 0}=-\frac{1}{2}\bar\eth \ln P, \nonumber\\
\gamma ^{ 0}&=& -\frac{\dot P }{2P}, \qquad \nu ^{ 0}=-2\bar \eth\gamma^{ 0}, \nonumber\\
\omega ^{ 0}&=&-\bar{\eth }\sigma ^{ 0}, \qquad \lambda ^{0}=\dot{\bar{\sigma }}^{0}-\bar {\sigma}^{ 0} \frac{\dot P}{P}, \\
\mu ^{0}&=&U^{ 0}=-\eth \bar{\eth} \ln P, \nonumber\\
\psi _{2}^{0}-\bar{\psi }_{2}^{0}&=&\bar{\eth }^{ 2}\sigma ^{ 0}-\eth^{2}\bar{\sigma }^{0}+\bar{\sigma }^{0}\lambda ^{0}-\sigma ^{0}\bar{\lambda }^{ 0}, \nonumber\\
\psi _{3}^{0}&=&\bar{\eth} \eth \bar{\eth} \ln P +\eth\lambda^{0}, \nonumber\\
\psi _{4}^{0}&=&-\bar\eth^{2} \left(\frac{\dot P}{P}\right)-\dot \lambda ^{0}+\lambda^{0} \frac{\dot P}{P}. \nonumber
\end{eqnarray}
Finally, in the NP formalism the Bianchi identities tell us about the time evolution of the asymptotic fields,
\begin{eqnarray}
\dot{\psi}_{2}^{0}-3\psi _{2}^{0}\frac{\dot{P}}{P} &=&-\eth \psi
_{3}^{0}+\sigma ^{0}\psi _{4}^{0}+2\phi _{2}^{0}\bar{\phi}_{2}^{0}, \nonumber\\
\dot{\psi}_{1}^{0}-3\psi _{1}^{0}\frac{\dot{P}}{P} &=&-\eth \psi
_{2}^{0}+2\sigma ^{0}\psi _{3}^{0}+4\phi _{1}^{0}\bar{\phi}_{2}^{0} \nonumber\\
\dot{\psi}_{0}^{0}-3\psi _{0}^{0}\frac{\dot{P}}{P} &=&-\eth\psi
_{1}^{0}+3\sigma ^{0}\psi _{2}^{0}+6\phi _{2}^{0}\bar{\phi}_{2}^{0}, \\
\dot{\phi}_{1}^{0}-2\phi _{1}^{0}\frac{\dot{P}}{P} &=&-\eth \phi _{2}^{0} ,\nonumber\\
\dot{\phi}_{0}^{0}-2\phi _{0}^{0}\frac{\dot{P}}{P} &=&-\eth \phi
_{1}^{0}+\sigma ^{0}\phi _{2}^{0}, \nonumber
\end{eqnarray}
here the dots represent $\partial_u$, and $\eth, \bar\eth$ are the differential operators defined by eqs. (\ref{eth}) and (\ref{ethb}).

\section{Bondi coordinates} \label{sec1.5}
As mentioned in Section \ref{sec1.1}, it is possible to introduce a general set of coordinates in the neighborhood of  $\scri^+$ since the conformal factor $P$, in eq. (\ref{metrica}), provides a great freedom when they are chosen. However, in many practical applications, it is useful to restrict the transformation imposing the condition
\begin{equation}\label{Pbondi}
P=P_0=1+\zeta \bar\zeta.
\end{equation}
With this choice, the two-surface metric (\ref{metrica}) becomes a sphere. These coordinates are then called Bondi coordinates. In this section, we will give the main equations of the NP formalism written in the Bondi system. These particular systems will correspond to inertial frames in General Relativity. However, the choice of the Bondi coordinate system is not unique, the coordinate transformations between two Bondi systems is called the Bondi-Metzner-Sachs (BMS) transformations \cite{BMS}. Now, making the choice of the $P$ factor by imposing eq. (\ref{Pbondi}), and since $\dot P=0$, one can reduce the equations introduced in section \ref{SC} as follows
\begin{eqnarray}
\psi _{2}^{0}-\bar{\psi }_{2}^{0}&=&\bar{\eth }^{2}\sigma ^{0}-\eth^{2}\bar{\sigma }^{0}+\bar{\sigma }^{0}\dot\sigma ^{0}-\sigma ^{0}\dot{\bar{\sigma}} ^0, \nonumber\\
\psi _{3}^{0}&=&\eth \dot{\bar{\sigma}} ^0, \label{radiation}\\
\psi _{4}^{0}&=&-\ddot{\bar{\sigma}} ^0. \nonumber
\end{eqnarray}
Also the ``eth'' operators $\eth,\bar\eth$ can be written as
\begin{eqnarray}
\eth f&=&P_0^{1-s}\frac{\partial (P_0^{s}f)}{\partial \zeta }, \label{eth}\\
\bar{\eth }f&=&P_0^{1+s}\frac{\partial (P_0^{-s}f)}{\partial \bar{\zeta}}. \label{ethb}
\end{eqnarray}
In many applications, it is quite convenient to define the so called Mass Aspect $\Psi$ from the Weyl scalar $\psi^0_2$ \cite{BBM},
\begin{equation}\label{asp.masa}
\Psi=\psi _{2}^{0}+\eth^{2}\bar{\sigma }^{0}+\sigma ^{0}\dot{\bar{\sigma}}^0,
\end{equation}
which satisfies the reality condition $\Psi=\bar\Psi$. Finally the Bianchi identities in the Bondi system take the form,
\begin{eqnarray}
\dot{\psi}_{2}^{0} &=&-\eth \psi_{3}^{0}+\sigma ^{0}\psi _{4}^{0}+2\phi _{2}^{0}\bar{\phi}_{2}^{0}, \label{psi2dot} \\
\dot{\psi}_{1}^{0} &=&-\eth \psi_{2}^{0}+2\sigma ^{0}\psi _{3}^{0}+4\phi _{1}^{0}\bar{\phi}_{2}^{0}, \label{psi1dot}\\
\dot{\psi}_{0}^{0} &=&-\eth \psi_{1}^{0}+3\sigma ^{0}\psi _{2}^{0}+6\phi _{2}^{0}\bar{\phi}_{2}^{0}, \\
\dot{\phi}_{1}^{0} &=&-\eth \phi _{2}^{0}, \label{phi1dot}\\
\dot{\phi}_{0}^{0} &=&-\eth \phi_{1}^{0}+\sigma ^{0}\phi _{2}^{0}. \label{phi0dot}
\end{eqnarray}
Also, it is possible to write eq. (\ref{psi2dot}) in terms of $\Psi$ as follows,
\begin{equation}\label{asp.masadot}
\dot\Psi=\dot\sigma^0\dot{\bar{\sigma}}^0+2\phi _{2}^{0}\bar{\phi}_{2}^{0}.
\end{equation}
In the same way, the SC can be written in terms of the shear $\sigma^0$ and the Weyl scalars $\psi^0_n$.

\subsection{The Mass and Bondi momentum}
In a Bondi system, it is possible to define the four-momentum vector for an asymptotically flat spacetime as \cite{nu},
\begin{equation}
P^a=-\frac{1}{8\pi \sqrt{2}}\int_S \Psi\tilde{l}^{a}d\Omega,
\end{equation}
where the vector $\tilde{l}^{a}$ is given by
\begin{equation}
\tilde{l}^{a}=\frac{1}{1+\zeta \bar{\zeta}}(1+\zeta \bar{\zeta},\zeta +\bar{\zeta}%
,-i(\zeta -\bar{\zeta}),1-\zeta \bar{\zeta}),
\end{equation}
and where $d\Omega=\frac{4id\zeta \wedge d\bar\zeta}{P_0^2}$ is the area of the unit sphere. Immediately from this definition it follows that the Bondi mass can be written as
\begin{equation}\label{massB}
M =-\frac{c^{2}}{8\pi \sqrt{2}G}\int_S \Psi d\Omega.
\end{equation}
The Bondi mass agrees with the usual definition of mass, e.g. with the Schwarzschild mass, and it is positive in a neighborhood of $\scri^+$.
Now, taking the time derivative of (\ref{massB}), and using eq. (\ref{asp.masadot}), we have
\begin{equation}
\dot M =-\frac{c^{2}}{8\pi \sqrt{2}G}\int_S (\dot\sigma^0\dot{\bar{\sigma}}^0+2\phi _{2}^{0}\bar{\phi}_{2}^{0}) d\Omega,
\end{equation}
since the integral is always positive, the r.h.s of the last equation is negative, i.e.
\begin{equation}
\dot M < 0 \qquad if \qquad \dot\sigma^0 \neq 0.
\end{equation}
Thus $\dot M$ measures the amount of mass loss carried away as gravitational radiation. Note that, in the astrophysical systems of major interest for the gravitational wave observatories like LIGO, the contribution of the electromagnetic radiation is several orders of magnitude less than the gravitational one, e.g. in an astrophysical process such as binary coalescence.

\section{The BMS group} \label{sec.4}
The set of coordinate transformations at $\scri^+$ preserving the conditions in the Bondi coordinates is called the Bondi-Metzner-Sachs Group (BMS) \cite{sachs,BMS,sachs2}. This group is the same as the asymptotic symmetry group that arises from the infinitesimal generators, i.e. from the asymptotic Killing vectors. Now, to construct the BMS group, we start considering the following mapping
\begin{eqnarray}
\hat{u}&=&T(u,\zeta,\bar \zeta),\label{nuBMS0}\\
\hat{r}&=& \dot T^{-1}r ,\label{nrBMS}\\
\hat{\zeta} &=& \frac{a\zeta+b}{c\zeta+d} \qquad ad-bc=1, \label{anguloBMS}
\end{eqnarray}
where $u,\zeta,$ and $r$ are the standard Bondi-type coordinates. Under this coordinate transformation all the relations developed up to this point are preserved. The fraction linear transformations (\ref{anguloBMS}) are the only one-to-one mapping of the sphere to itself. Then, the BMS group is defined by the mapping (\ref{anguloBMS}) and the following restriction on the transformations (\ref{nuBMS0}),
\begin{equation}
\hat{u}=K(u+\alpha), \label{nuBMS}
\end{equation}
where $\alpha(\zeta,\bar\zeta)$ is a regular arbitrary function on the sphere. Now, it is possible to show from eq. (\ref{nuBMS}) and (\ref{anguloBMS})
that the spherical metric transforms as \cite{HNP},
\begin{equation}
\frac{4 d\hat{\zeta} d\bar{\hat\zeta}}{\hat{P}_0^{2}}=K^2\frac{4d\zeta d\bar\zeta}{P_0^2},
\end{equation}
where the conformal factor $K$ associated with this transformation is given by
\begin{eqnarray}
K&=& J^{-\frac{1}{2}} \frac{P_{0}}{\hat{P}_0}, \\
&=& (1+\zeta \bar\zeta)[(a\zeta +b)(\bar a+ \bar \zeta+ \bar b)+(c\zeta+d)(\bar c \bar \zeta+ \bar d)], \nonumber
\end{eqnarray}
and where
\begin{eqnarray}
J&=&\frac{\partial \zeta}{\partial \hat{\zeta}}\frac{\partial \bar \zeta}{\partial \bar {\hat{\zeta}}}, \\
P_{0}&=& 1+\zeta \bar\zeta,\\
\hat{P}_0&=& 1+\hat{\zeta} \bar{\hat{\zeta}}.
\end{eqnarray}
Now, the infinite-parameter subgroup obtained by setting $K=1$,
\begin{eqnarray}
\hat{\zeta}&=&\zeta, \label{zetaigual}\\
\hat{u}&=&u+\alpha, \label{stra}
\end{eqnarray}
is known as the supertranslation subgroup. A supertranslation $\alpha(\zeta,\bar\zeta)$ moves points on each generator by an amount $\alpha(\zeta,\bar\zeta)$. This function can be expressed in terms of infinite constants using, for example, a tensorial spin-s harmonic expansion \cite{ngilb} in the following way,
\begin{equation}
\alpha=\alpha^0+\alpha^iY_{1i}^0(\zeta,\bar\zeta)+\alpha^{ij}Y_{2ij}^0(\zeta,\bar\zeta)+...
\end{equation}
In this expansion, $\alpha^0$, and $\alpha^i$ represents the ordinary translations. For extra details about the Lorentz group and the BMS transformations the reader could see refs. \cite{HNP,newman1966note}.

\subsection{Transformation between systems}
In this subsection, we discuss the transformation laws between two frames. For that, we assume that the coordinates of these frames are related by eqs. (\ref{nrBMS}), (\ref{nuBMS}), and (\ref{zetaigual}), i.e by the following mapping,
\begin{eqnarray}
\hat{u}&=&T(u,\zeta,\bar \zeta), \\
\hat{r}&=& \dot T^{-1}r, \\
\hat{\zeta} &=& \zeta.
\end{eqnarray}
Now, for this set of coordinates $(\hat{u},\hat{r},\zeta,\bar\zeta)$, one can build a new tetrad system just following the steps of sec. \ref{sec1.1}, where the first null vector is chosen as $\hat l_a=\nabla_a \hat{u}$. Then, the remaining vectors are chosen to satisfy the orthonormality conditions. So, we get a new null tetrad, these vectors are labeled as $(\hat{l}^a,\hat{n}^a,\hat{m}^a,\bar{\hat{m}}^a)$.  Also, one can find the relation between the null basis $(l^a,{n}^a,{m}^a,\bar{m}^a)$ obtained from the coordinates $(u,r,\zeta,\bar\zeta)$ with the previous one given by eqs. (\ref{la}-\ref{na}). These equations can be expressed in the following way \cite{KQ2},
\begin{eqnarray}
\hat{l}^{a} &=& \dot T \left( l_{a}-\frac{L}{r}\bar{m}_{a}-\frac{\bar{L}}{r}m_{a}+\frac{L \bar{L}}{r^2}n_{a}\right), \label{trans1} \\
\hat{n}^{a} &=& \dot T^{-1} n^{a}, \\
\hat{m}^{a} &=& m^{a}-\frac{L}{r} n^{a}, \\
\bar{\hat{m}}^{a} &=& \bar{m}^{a}-\frac{\bar{L}}{r} n^{a}, \label{trans2}
\end{eqnarray}
where
\begin{equation}
L=-\frac{\eth_{(u)} T}{\dot T},
\end{equation}
here $\eth_{(u)}$ means applying the $\eth$ operator keeping $u$ as a constant. Now, using the set of equations (\ref{trans1}-\ref{trans2}), it is possible to expand the scalars defined in the new system in terms of the scalars defined in the original frame. As an example, we start with $\hat{\psi}_1$,
\begin{eqnarray*}
\hat{\psi}_{1} &=&-{C_{abc}}^{d}\hat{l}^{a}\hat{n}^{b}\hat{l}^{c}\hat{m}_{d}, \\
&=&\dot{T}[{\psi }_{1}-3\frac{L}{r}\psi
_{2}+3\frac{L^{2}}{r^{2}}\psi _{3}-\frac{L^{3}}{r^{3}}\psi _{4}].
\end{eqnarray*}
Finally, assuming  the ``peeling'' and using eq. (\ref{nrBMS}) we can write,
\begin{equation}\label{transPsi}
\hat{\psi}_{1}^{0}= \dot{T}^{-3}[{\psi }_{1}^{0}-3L\psi
_{2}^{0}+3L^{2}\psi _{3}^{0}-L^{3}\psi _{4}^{0}].
\end{equation}
In the same way, we can find the transformation law for all Weyl scalars, Maxwell scalars, and the spin coefficients (particularly the shear), which can be listed as follows \cite{KQ2},
\\
\emph{Weyl scalars:}
\begin{eqnarray}
\hat{\psi}_{0}^{0}&=&\dot{T}^{-3}[\psi _{0}^{0}-4L\psi
_{1}^{0}+6L^{2}\psi _{2}^{0}-4L^{3}\psi _{3}^{0}+L^{4}\psi _{4}^{0}] ,\nonumber\\
\hat{\psi}_{1}^{0}&=&\dot{T}^{-3}[{\psi }_{1}^{0}-3L\psi
_{2}^{0}+3L^{2}\psi _{3}^{0}-L^{3}\psi _{4}^{0}], \nonumber\\
\hat{\psi}_{2}^{0}&=&\dot{T}^{-3}[\psi _{2}^{0}-2L\psi
_{3}^{0}+L^{2}\psi _{4}^{0}], \\
\hat{\psi}_{3}^{0}&=&\dot{T}^{-3}[\psi _{3}^{0}-L\psi _{4}^{0}], \nonumber\\
\hat{\psi}_{4}^{0} &=&\dot{T}^{-3}\psi _{4}^{0}. \nonumber
\end{eqnarray}
\emph{Maxwell scalars:}
\begin{eqnarray}
\hat{\phi}_{0}^{0}&=&\dot{T}^{-2}[\phi _{0}^{0}-2L\phi
_{1}^{0}+L^{2}\phi _{2}^{0}], \nonumber\\
\hat{\phi}_{1}^{0} &=&\dot{T}^{-2}[\phi _{1}^{0}-L\phi _{2}^{0}], \\
\hat{\phi}_{2}^{0} &=&\dot{T}^{-2}\phi _{2}^{0}. \nonumber
\end{eqnarray}
\emph{Shear:}
\begin{equation}
\hat{\sigma}^{0}=\dot{T}^{-1}[\sigma ^{0}-\eth_{(u)} L- L \dot L].
\end{equation}

\section{Physical interpretation of the asymptotic scalars} \label{sec.5}
In this section, we will discuss the physical interpretation of some of the complex scalars used in the NP formalism. A simpler way to introduce this topic, is to focus on the dominant terms of the peeling expansion, and make a tensorial spin-s harmonic expansion of these functions. The most important scalars in our analysis are the following,
\begin{eqnarray}
\psi _{1}^{0} &=&\psi _{1}^{0i}(u)Y_{1i}^{1}(\zeta,\bar \zeta )+\psi_{1}^{0ij}(u)Y_{2ij}^{1}(\zeta,\bar \zeta ),\nonumber \\
\psi _{2}^{0}  &=&\psi_2^{00}+\psi_2^{0i}Y_{1i}^{0}(\zeta,\bar \zeta )+\psi_2^{0ij}(u)Y_{2ij}^{0}(\zeta,\bar \zeta ),\nonumber\\
\sigma^0 &=&\sigma^{ij}(u)Y_{2ij}^{2}(\zeta,\bar \zeta ),\nonumber\\
\phi _{0}^{0} &=&\phi _{0}^{0i}(u)Y_{1i}^{1}(\zeta,\bar \zeta ), \label{exp} \\
\phi _{1}^{0} &=&\phi _{1}^{00}+\phi _{1}^{0i}(u)Y_{1i}^{0}(\zeta,\bar \zeta ).\nonumber
\end{eqnarray}
For any asymptotically flat axially symmetric spacetime, the Komar integral \cite{komar} gives a precise notion of the angular momentum. The Komar formula uses the Killing vector field to define the global angular momentum. Assuming that the axis of symmetry is labeled as z-axis, the non-zero component of the angular momentum can be written as follows \cite{KQ},
\begin{equation}
Im[\psi_1^0-\sigma^0\eth\bar\sigma^0]^z=-\frac{6\sqrt{2}G}{c^3}J^z.
\end{equation}
Since the real contribution of $[\sigma^0\eth\bar\sigma^0]^i$ is zero for any axisymmetric spacetimes. One only need to use the real part of $\psi _{1}^{0i}$ to define the dipole mass moment as follows,
\begin{equation}
Re[\psi_1^0]^z=-\frac{6\sqrt{2}G}{c^3}D^z.
\end{equation}
Now, if the spacetime has no global symmetries, the mass dipole-angular momentum two form will add some extra contributions of the free data $\sigma^0$ and its derivatives. Notably, as one can see in the literature, there are many definitions of the angular momentum for isolated systems in general relativity. A recent living review \cite{Szab} offers a complete survey of the main results in the field with the main motivations and technical aspects of each definition, the fact that there is no agreement among these alternative approaches reflects the difficulty of the subject. Although, in a recent work together with Kozameh \cite{KQ2}, we introduced a new definition of dipole mass moment-angular momentum tensor using the Winicour-Tamburino linkage \cite{TW},
\begin{equation}\label{DJB}
\left[ 2\psi _{1}^{0}-2\sigma^{0}\eth \bar{\sigma}^{0}-\eth(\sigma ^{0}\bar{\sigma}^{0})\right]^i=-\frac{12\sqrt{2}G}{c^{2}}[D^{i}+ic^{-1}J^{i}].
\end{equation}
This definition allows to define the center of mass and write the equation of motion of the center of mass linking the time evolution with the emitted gravitational radiation.
On the other hand, as we discuss in sec. \ref{sec1.5}, for a Bondi system, there is a precise notion of mass and linear momentum. Now, we can use the $\ell=0,1$ components of the tensorial expansion previously introduced, and relate $\psi _{2}^{0}$ to the Bondi 4-momentum $(M,P^i)$, as shown in the following equations,
\begin{eqnarray}
\left[\psi _{2}^{0}+\eth^{2}\bar{\sigma }^{0}+\sigma ^{0}\dot{\bar{\sigma}}^0\right]|_{\ell=0}&=&-\frac{2\sqrt{2}G}{c^{2}}M, \label{massM}\\
\left[\psi _{2}^{0}+\eth^{2}\bar{\sigma }^{0}+\sigma ^{0}\dot{\bar{\sigma}}^0\right]^i&=&-\frac{6G}{c^{3}}P^{i}. \label{momentoP}
\end{eqnarray}
The $\ell\geq 2$ terms of the l.h.s of eqs. (\ref{massM}) or (\ref{momentoP}), are the so called ``supermomentum'' at null infinity.
Now, in the NP approach, the scalar $\Psi_4$ measures the gravitational radiation received at null infinity. This scalar is related to the gravitational wave modes as follows,
\begin{equation}
\Psi_4=\ddot{h}_{+}-i\ddot{h}_{\times},
\end{equation}
where $h_{+},$ $h_{\times}$ are the plus mode, and cross mode of the gravitational wave in the transverse traceless gauge \cite{ttgauge}, respectively. Thus, the complex function $\ddot{\sigma}^0$, introduced in eq. (\ref{radiation}), yields the gravitational radiation reaching at null infinity, and ${\sigma}_R^{ij}=h_{+}^{ij}$, and ${\sigma}_I^{ij}=h_{\times}^{ij}$ are the quadrupolar contributions of the gravitational wave.

Finally, we focus on the Maxwell scalars, which give information about the electromagnetic contribution received at $\scri^+$. At this point, we can mention that the electric charge is the zeroth order in the tensorial expansion of $\phi _{1}^{0}$, and the dipole electromagnetic moment corresponds to the $\ell=1$ component of $\phi _{0}^{0}$ \cite{quiroga2017sources,kozameh2005electromagnetic}, i.e,
\begin{eqnarray}
\phi _{1}^{00}&=&Q, \\
\frac{1}{2}\phi _{0}^{0i}&=&p^{i}+ic^{-1}\mu^{i},
\end{eqnarray}
where $p^i,\mu^i$ are the electric and magnetic dipole moment respectively.

\section{Application to stationary axisymmetric spacetimes} \label{sec:2}

Stationary axisymmetric spacetimes are of great importance in General Relativity, astrophysics, Newtonian gravity, and also in Post-Newtonian theories. Many sources like stars, galaxies, accretion disks, and black holes are modeled under these assumptions of temporal and axial symmetry. These global symmetries play a central role in analytic calculations, since they are very useful when the field equations are simplified. In this section, we will focus on studying these kinds of spacetimes. Now, following ref. \cite{chandrasekhar1998mathematical}, we can introduce a more general stationary axisymmetric metric in the standard spherical coordinates  $(t,r,\theta ,\varphi )$ as follows,
\begin{align}
ds^{2}=&e^{2\mu _{0}}dt^{2}-e^{2\mu _{1}}dr^{2}-r^{2}e^{2\mu _{2}}d\theta^{2} \nonumber\\
&-r^{2}e^{2\mu _{3}}\sin ^{2}\theta (d\varphi -\omega dt)^{2}, \label{gabS}
\end{align}
where the metric functions $\mu _{0}$, $\mu _{1}$, $\mu _{2}$, $\mu _{3}$, and $\omega$ are arbitrary functions of $(r,\theta)$. The stationary and axisymmetric character of (\ref{gabS}), is reflected in the fact that the metric coefficients are independent of the coordinate $t$, and the azimuthal angle $\varphi$, and also that the spacetime is invariant under simultaneous transformations $t \rightarrow -t$ and $\varphi \rightarrow -\varphi$. Now, we are considering spacetimes which are asymptotically flat in the neighborhood of infinity, thus to ensure an adequate behaviour of the line element (\ref{gabS}), the metric functions $\mu _{0}$, $\mu _{1}$, $\mu _{2}$, $\mu _{3}$, and the angular velocity $\omega$ must go to zero as $r\rightarrow \infty$. Also we choose the signature $(+,-,-,-)$ in agreement with the orthonormal conditions introduced in Sec. \ref{sec1.1}.

One of the main ingredients in the NP formalism is the construction of a complex null tetrad $(l^a, n^a, m^a, \bar{m}^a)$. For that, we introduce first an orthogonal tetrad constructed from the timelike foliation $t=const$, then the normal vector to the hypersurface $\Sigma_t$ is given by
\begin{equation}
t^a=[e^{-\mu_0},0,0,\omega e^{-\mu_0}].
\end{equation}
Now, on $\Sigma_t$ we can find three spacelike vectors denoted by
\begin{align}
r^a&=[0,-e^{-\mu_1},0,0],\\
e_{\theta}^a&=[0,0,-\frac{e^{-\mu_2}}{r},0],\\
e_{\varphi}^a&=[0,0,0,-\frac{e^{-\mu_3}}{r \sin \theta}].
\end{align}
Also, the set of vectors $(t^a,r^a,e_{\theta}^a,e_{\varphi}^a)$ satisfy
\begin{equation}
g_{ab}t^at^b=-g_{ab}r^ar^b=-g_{ab}e_{\theta}^ae_{\theta}^b=-g_{ab}e_{\varphi}^ae_{\varphi}^b=1.
\end{equation}
Now, from these vectors we can build a null tetrad making the following linear combinations,
\begin{align}
l^a&=\frac{1}{\sqrt{2}}(t^a+r^a), \\
n^a&=\frac{1}{\sqrt{2}}(t^a-r^a),\\
m^a&=\frac{1}{\sqrt{2}}(e_{\theta}^a-ie_{\varphi}^a),\\
\bar{m}^a&=\frac{1}{\sqrt{2}}(e_{\theta}^a+ie_{\varphi}^a),
\end{align}
then, our null tetrad is given by the null vectors,
\begin{align}
l^{a} &=[\frac{e^{-\mu _{0}}}{\sqrt{2}},-\frac{e^{-\mu _{1}}}{\sqrt{2}},0,%
\frac{\omega e^{-\mu _{0}}}{\sqrt{2}}], \label{laG}\\
n^{a} &=[\frac{e^{-\mu _{0}}}{\sqrt{2}},\frac{e^{-\mu _{1}}}{\sqrt{2}},0,%
\frac{\omega e^{-\mu _{0}}}{\sqrt{2}}], \\
m^{a} &=[0,0,-\frac{e^{-\mu _{2}}}{r\sqrt{2}},\frac{ie^{-\mu _{3}}}{\sqrt{2}%
r\sin \theta }], \\
\bar{m}^{a} &=[0,0,-\frac{e^{-\mu _{2}}}{r\sqrt{2}},\frac{-ie^{-\mu _{3}}}{%
\sqrt{2}r\sin \theta }]\label{mbaG},
\end{align}
and lowering the indices using $g_{ab}$ we find the conjugate tetrad which is given by
\begin{align}
l_{a} &=[\frac{1}{\sqrt{2}}e^{\mu _{0}},\frac{1}{\sqrt{2}}e^{\mu _{1}},0,0],
\\
n_{a} &=[\frac{1}{\sqrt{2}}e^{\mu _{0}},-\frac{1}{\sqrt{2}}e^{\mu _{1}},0,0],
\\
m_{a} &=[\frac{i}{\sqrt{2}}r\omega \sin \theta e^{\mu _{3}},0,\frac{1}{%
\sqrt{2}}re^{\mu _{2}},-\frac{i}{\sqrt{2}}r\sin \theta e^{\mu _{3}}], \\
\bar{m}_{a} &=[-\frac{i}{\sqrt{2}}r\omega \sin \theta e^{\mu _{3}},0,\frac{1%
}{\sqrt{2}}re^{\mu _{2}},\frac{i}{\sqrt{2}}r\sin \theta e^{\mu _{3}}].
\end{align}
Finally, we will assume the following potential vector
\begin{equation}
A_{a}=[\chi ,-A_{r},-A_{\theta },-A_{\varphi}],
\end{equation}
where $\chi$, $A_{r}$, $A_{\theta }$, and $A_{\varphi }$ are also functions of $(r,\theta)$. Now, using the null tetrad, and the potential vector $A_a$, we can compute all the quantities introduced in the previous sections to solve this axially symmetric stationary metric in the NP formalism. In the appendix A, we will show the general set of equations, but in the next three subsections, we will reduce the metric given by eq. (\ref{gabS}) to the Reissner-Nordström, Chazy-Curzon, and Kerr spacetimes, by solving an algebraic system of five equations for the metric functions of (\ref{gabS}), and we will write all the complex scalars on the basis given by eqs. (\ref{laG}-\ref{mbaG}).

\subsection{Reissner-Nordström spacetime}\label{sec:4:1}
The Reissner–Nordström metric is a static solution of the Einstein-Maxwell equations. This solution corresponds to the gravitational field of a charged, non-rotating, spherically symmetric body of mass $M$ and electric charge $Q$. The line element of this spacetime in spherical coordinates can be written as
\begin{align}
ds^2 =&\left( 1 - \frac{2M}{r} + \frac{Q^2}{r^2} \right) dt^2 -\left( 1 - \frac{2M}{r} + \frac{Q^2}{r^2} \right)^{-1} dr^2 \nonumber\\
&- r^2 (d\theta^2 + \sin^2\theta d\varphi^2).
\end{align}
Comparing to eq. (\ref{gabS}), and solving for the metric functions, we find the following algebraic equations
\begin{align}
\mu _{0} &=\frac{1}{2}\ln \left( 1-2\,{\frac{M}{r}}+{\frac{{Q}^{2}}{{r}%
^{2}}}\right), \\
\mu _{1} &=-\frac{1}{2}\ln \left( 1-2\,{\frac{M}{r}}+{\frac{{Q}^{2}}{{r}%
^{3}}}\right), \\
\mu _{2} &=\mu _{3}=\omega=0.
\end{align}
Finally, replacing in the equations of section \ref{sec:3} we can write,
\begin{align}
\phi _{0} &=\phi_{2}=0, \label{eq.1}\\
\psi _{0}=\psi_{1}&=\psi _{3}=\psi _{4}=0, \\
\sigma =\tau =\kappa &=\lambda =\nu =\pi =0,
\end{align}
and where
\begin{align}
\psi _{2} &=-{\frac{M}{{r}^{3}}}+O\left( {r}^{-4}\right), \\
\phi _{1} &=-{\frac{Q}{2{r}^{2}}}+O\left( {r}^{-4}\right), \\
\rho &=-{\frac{1}{\sqrt{2}r}}+O\left( {r}%
^{-2}\right), \\
\alpha &=-{\frac{\cot \theta }{2\sqrt{2}r}}, \\
\epsilon &={\frac{M}{2\sqrt{2}{r}^{2}}}+O\left( {r}^{-3}\right). \label{eq.2}
\end{align}
As we can see from eqs. (\ref{eq.1}) to (\ref{eq.2}), this exact solution is a shear free solution since $\sigma=0$.

\subsection{Chazy-Curzon solution}\label{sec:4:2}

The Chazy-Curzon solution corresponds to the simplest case of the Weyl vacuum solutions. Note that if we start from eq. (\ref{gabS}) and set $\mu_3=-\mu_0$, and $\mu_1=\mu_2=\gamma_0-\mu_0$, with $\gamma_0=\gamma_0(r,\theta)$,  we obtain the Weyl metric \cite{kramer1980exact} in spherical coordinates
\begin{align}
ds^{2}=&e^{2\mu _{0}}dt^{2}-e^{2\gamma_0-2\mu_0}(dr^{2}+r^{2}d\theta^{2})\nonumber\\
&-r^{2}e^{-2\mu _{0}}\sin ^{2}\theta d\varphi ^{2},
\end{align}
where the Chazy-Curzon spacetime is obtained by setting,
\begin{eqnarray*}
\mu_{0} &=&-\frac{M}{r}, \\
\mu_{1} &=&\mu_{2}=-\frac{M^{2}\sin ^{2}\theta }{2r^{2}}+\frac{M}{r}, \\
\mu_{3} &=&\frac{M}{r}.
\end{eqnarray*}
Now, the Weyl scalars can be written in the following way,
\begin{eqnarray*}
\psi _{0} &=&\,{\frac{{M}^{3}\left( \sin \theta \right) ^{2}}{2{r}^{5}}}%
+O\left( {r}^{-6}\right), \\
\psi _{1} &=&-\,{\frac{{M}^{3}\sin \theta \cos \theta }{2{r}^{5}}}+O\left( {r%
}^{-6}\right), \\
\psi _{2} &=&-{\frac{M}{{r}^{3}}}+O\left( {r}^{-4}\right), \\
\psi _{3} &=&{\frac{{M}^{3}\sin \theta \cos \theta }{2{r}^{5}}}+O\left( {r}%
^{-6}\right), \\
\psi _{4} &=&\,{\frac{{M}^{3}\left( \sin \theta \right) ^{2}}{2{r}^{5}}}%
+O\left( {r}^{-6}\right).
\end{eqnarray*}
Finally, the spin coefficients are given by,
\begin{eqnarray*}
\sigma &=&-\,{\frac{\sqrt{2}{M}^{2}\left( \sin \theta \right) ^{2}}{4{r}^{3}}%
}+O\left( {r}^{-4}\right), \\
\rho &=&-{\frac{\sqrt{2}}{2r}}+O\left( {r}^{-2}\right), \\
\tau &=&\,{\frac{\sqrt{2}{M}^{2}\sin \theta \cos \theta }{4{r}^{3}}}+O\left(
{r}^{-4}\right), \\
\kappa &=&-\,{\frac{\sqrt{2}{M}^{2}\sin \theta \cos \theta }{4{r}^{3}}}%
+O\left( {r}^{-4}\right), \\
\alpha &=&-{\frac{\sqrt{2}\cot \theta }{4r}}+O\left( {r}^{-2}\right), \\
\epsilon &=&\,{\frac{M\sqrt{2}}{4{r}^{2}}}+O\left( {r}^{-3}\right).
\end{eqnarray*}
Since Chazy-Curzon is a vacuum non-charged solution of the Weyl metric, it is clear that the Maxwell scalars will be zero.

\subsection{Kerr spacetime}\label{sec:4:3}
As a last example, we choose the Kerr metric which describes the geometry of a spacetime in the vicinity of a rotating mass $M$ with angular momentum $J$. This metric corresponds to a vacuum non-charged solution of the Einstein equations. In usual coordinates, the Kerr line element can be written as,
\begin{align}
ds^2&= \left( 1 - \frac{2Mr}{{\hat{\rho}}^{2}} \right)dt^{2} - \frac{\hat{\rho}^{2}}{\Delta} dr^{2} - \rho^{2} d\theta^{2} \nonumber\\
&+ \frac{4Mr a\sin^{2}\theta }{\hat{\rho}^{2}}dt d\varphi  \nonumber\\
& -\left( r^{2} + a^{2} + \frac{2Mr a^{2}}{\hat{\rho}^{2}} \sin^{2} \theta \right) \sin^{2} \theta \ d\varphi^{2},
\end{align}
where $\hat{\rho},\Delta,$ and $a$ have been introduced for brevity, these functions are given by
\begin{eqnarray*}
{\hat{\rho}}^{2} &=&{r}^{2}+{a}^{2}\cos ^{2}\theta, \\
\Delta &=&r^{2}-2Mr+a^{2}, \\
a &=&\frac{J}{M}.
\end{eqnarray*}
Solving for $\mu_0,\mu_1,\mu_2,\mu_3$, and $\omega$, we find
\begin{eqnarray*}
\mu _{0} &=&\frac{1}{2}\ln \left( {\frac{2\,Mr({a}^{2}\sin ^{2}\theta -{a}%
^{2}-{r}^{2})+({a}^{2}+{r}^{2}){\hat{\rho}}^{2}}{2\,Mr{a}^{2}\sin ^{2}\theta
+({a}^{2}+{r}^{2}){\hat{\rho}}^{2}}}\right),  \\
\mu _{1} &=&\frac{1}{2}\ln \left( {\frac{{\hat{\rho}}^{2}}{\Delta }}\right),
\\
\mu _{2} &=&\frac{1}{2}\ln \left( {\frac{{\hat{\rho}}^{2}}{r^{2}}}\right),  \\
\mu _{3} &=&\frac{1}{2}\ln \left( {\frac{2\,Mr{a}^{2}\sin ^{2}\theta +({a}%
^{2}+{r}^{2}){\hat{\rho}}^{2}}{{r}^{2}{\hat{\rho}}^{2}}}\right),  \\
\omega  &=&{\frac{2Mra}{2\,Mr{a}^{2}\sin ^{2}\theta +({a}^{2}+{r}^{2}){\hat{%
\rho}}^{2}}}.
\end{eqnarray*}
Finally, the Weyl scalars are given by,
\begin{eqnarray*}
\psi _{0} &=&\frac{3}{2}\,{\frac{{J}^{2}\sin ^{2}\theta }{{Mr}^{5}}}+O\left(
{r}^{-6}\right), \\
\psi _{1} &=&\frac{3i}{2}{\frac{J\sin \theta }{{r}^{4}}}+O\left( {r}%
^{-5}\right), \\
\psi _{2} &=&-{\frac{M}{{r}^{3}}}+O\left( {r}^{-4}\right), \\
\psi _{3} &=&-\frac{3i}{2}{\frac{J\sin \left( \theta \right) }{{r}^{4}}}%
+O\left( {r}^{-5}\right), \\
\psi _{4} &=&\frac{3}{2}{\frac{{J}^{2}\sin ^{2}\theta }{{Mr}^{5}}}+O\left(
{r}^{-6}\right),
\end{eqnarray*}
and the asymptotic solution of the spin coefficients are the following,
\begin{eqnarray*}
\sigma  &=&\frac{\sqrt{2}}{4}\,{\frac{{J}^{2}\sin ^{2}\theta }{{M}^{2}{r}^{3}%
}}+O\left( {r}^{-4}\right),  \\
\rho  &=&\frac{1}{\sqrt{2}r}\,+O\left( {r}^{-2}\right),  \\
\kappa  &=&\frac{\sqrt{2}}{4}\,{\frac{{J}^{2}\sin \theta \cos \theta }{M^{2}{%
r}^{3}}}+O\left( {r}^{-4}\right),  \\
\tau  &=&-\frac{\sqrt{2}}{4}\,{\frac{J\left( J\cos \theta
-6\,i{M}^{2}\right) \sin \left( \theta \right) }{{M}^{2}{r}^{3}}}+O\left( {r}%
^{-4}\right),  \\
\alpha  &=&\frac{\sqrt{2}}{4}{\frac{\cot \theta }{r}}+O\left( {r}%
^{-3}\right),  \\
\epsilon  &=&-\frac{\sqrt{2}}{4}{\frac{M\,}{{r}^{2}}}+O\left( {r}^{-3}\right).
\end{eqnarray*}
Note that the angular momentum of the Kerr solution is given by the imaginary part of $\psi_1^0$ as we mention in sec. (\ref{sec.5}). Additionally, the Kerr spacetime has not gravitational wave since the Bondi free data is equal to zero, i.e. $\sigma^0=0$.

\section{Final remarks}
In this article we give a brief review of the Newman-Penrose formalism and the asymptotic structure of the spacetime. This  review  has  highlighted  the  importance  of the complex scalar functions introduced in the NP approach to describe the dynamics of compact sources and its gravitational radiations. We show how to compute the physical variables from the fields received at null infinity, and we deduce the transformation laws for systems defined by different families of null cuts.  Finally, as an example, we use the asymptotic formulation of the general relativity in a general stationary axisymmetric metric. For that, we introduce a complex null tetrad and we compute the Weyl scalars and spin coefficients, then we reduce this general spacetime to some familiar exact solutions.

\subsection*{Acknowledgements}
GDQ wants to thank the financial support of VIE-UIS and the postdoctoral research program RC N001-1518-2016.

\appendix
\section{Complex scalars and Spin Coefficients for a more general stationary axisymmetric spacetimes}\label{sec:3}
In the following appendix, we present a set of equations derived from the line element (\ref{gabS}). These equations correspond to the Weyl and Maxwell scalars, and the spin coefficients computed from the tetrad basis introduced in sec. \ref{sec:2}. Now, starting with the Weyl scalars we can write,
\begin{widetext}
\begin{eqnarray}
\psi _{0} &=&-\frac{\sin ^{2}\theta }{4}\,e{^{-2\,\mu _{0}-2\,\mu
_{2}+2\,\mu _{3}}}\left( {\frac{\partial \omega }{\partial \theta }}\right)
^{2}-\frac{{e{^{-2\,\mu _{1}}}}}{2r}\,\frac{\partial }{\partial r}{\left(
\,\mu _{2}-\,\mu 3\right) -}\frac{e{^{-2\,\mu _{1}}}}{4}\,\left( {\frac{%
\partial \mu _{0}}{\partial r}}\frac{\partial }{\partial r}\left( \mu
_{3}-\mu _{2}\right) +{\frac{\partial \mu _{3}}{\partial r}}\frac{\partial }{%
\partial r}\left( \mu _{1}-\mu _{3}\right) \right)  \nonumber\\
&&-\frac{e{^{-2\,\mu _{1}}}}{4}\left( \frac{\partial }{\partial r}\left( \mu
_{2}-\mu _{1}\right) {\frac{\partial \mu _{2}}{\partial r}}+{\frac{\partial
^{2}}{\partial {r}^{2}}(}\mu _{2}-\mu _{3})\right) -\frac{{e{^{-2\,\mu _{2}}}%
}}{4r^{2}}\,{\left( \frac{\partial }{\partial \theta }\left( \mu _{2}-\mu
_{0}+\mu _{3}\right) {\frac{\partial \mu _{0}}{\partial \theta }}+{\frac{%
\partial }{\partial \theta }}\left( \mu _{1}-\mu _{2}-\mu 3\right) {\frac{%
\partial \mu _{1}}{\partial \theta }}\right) }\nonumber \\
&&{-\frac{{e{^{-2\,\mu _{2}}}}}{4r^{2}}}\left( {{\frac{\partial ^{2}}{%
\partial {\theta }^{2}}(}\mu _{1}-\mu _{0})+\cot \theta \frac{\partial }{%
\partial \theta }\left( \mu _{0}-\mu _{1}\right) }\right) +\frac{i\sin
\theta }{4}\left( {\frac{\partial \omega }{\partial \theta }}\frac{\partial
}{\partial r}\left( \mu _{2}-3\mu _{3}+2\mu _{0}\right) -{\frac{\partial
^{2}\omega }{\partial \theta \partial r}}\right) e{^{-\mu _{0}-\,\mu
_{1}-\,\mu _{2}+\mu _{3}}} \nonumber\\
&&-\frac{i\sin \theta }{4}\left( {\frac{\partial \omega }{\partial r}}\frac{%
\partial }{\partial \theta }\left( \mu _{0}+\mu _{1}\right) +\frac{2}{r}{{%
\frac{\partial {\omega }}{\partial \theta }}}\right) e{^{-\mu _{0}-\,\mu
_{1}-\,\mu _{2}+\mu _{3}}},
\end{eqnarray}
\begin{eqnarray}
\psi _{1} &=&\frac{r\sin ^{2}\theta }{2}{\frac{\partial \omega }{\partial
\theta }\frac{\partial \omega }{\partial r}}e{^{-2\,\mu _{0}-\,\mu _{1}-\mu
_{2}+2\mu 3}}+\frac{ir\sin \theta }{8}\left( {\frac{\partial \omega }{%
\partial r}}\frac{\partial }{\partial r}(\mu _{2}+{9\,}\mu _{3}-3\mu
_{0}-3\mu _{1}{)}+3{\frac{\partial ^{2}\omega }{\partial {r}^{2}}}\right) e{%
^{-\,\mu _{0}-2\,\mu _{1}+\mu _{3}}}\nonumber \\
&&{+\frac{{e{^{-\,\mu _{1}-\mu _{2}}}}}{4r}\,\left( \frac{\partial }{%
\partial r}\left( \mu _{2}-\,\mu _{3}\right) {\frac{\partial \mu _{3}}{%
\partial \theta }}-\,{\frac{\partial ^{2}}{\partial \theta \partial r}(}\mu
_{3}+3\mu _{0})\right) }+\frac{3i{\cos \theta }}{8r}{{\frac{\partial {\omega
}}{\partial \theta }}e{^{-\mu _{0}-2\mu _{2}+\mu _{3}}}} \nonumber\\
&&+\frac{5i\sin \theta }{4}\,{\frac{\partial \omega }{\partial r}e{^{-\,\mu
_{0}-2\,\mu _{1}+\mu _{3}}}}+{\frac{i{\sin \theta }}{8r}}\left( {\frac{%
\partial }{\partial \theta }(3\mu _{1}-\mu _{0}+3\,\mu _{3}-\mu _{2})\frac{%
\partial \omega }{\partial \theta }+{\frac{\partial ^{2}{\omega }}{\partial {%
\theta }^{2}}}}\right) {e^{-\mu _{0}-2\mu _{2}+\mu _{3}}} \nonumber\\
&&+{\frac{{e{^{-\,\mu _{1}-\mu _{2}}}}}{4r}\left( \frac{1}{r}\frac{\partial
}{\partial \theta }{\left( 3\,\mu _{0}+\mu _{1}\right) +}3\frac{\partial }{%
\partial r}\left( \,\mu _{2}-\mu _{0}\right) \frac{\partial \mu _{0}}{%
\partial \theta }+\frac{\partial }{\partial r}\left( 3\,\mu _{0}+\mu
_{3}\right) {\frac{\partial \mu _{1}}{\partial \theta }-}4{\cot \theta \,%
\frac{\partial }{\partial r}\left( \mu _{3}-\mu _{2}\right) }\right) },\\
\psi _{2} &=&-\frac{{{r}^{2}\sin ^{2}\theta }}{12}\left( 7{\,\left( {\frac{%
\partial }{\partial r}}\omega \right) ^{2}}+\,\frac{1}{r^{2}}\left( {\frac{%
\partial }{\partial \theta }}\omega \right) ^{2}\right) e{^{-2\,\mu
_{0}-2\mu _{2}+2\mu _{3}}}+\frac{e{^{-2\,\mu _{1}}}}{12}\,\frac{\partial }{%
\partial r}\left( \mu _{2}-\,\mu _{1}+\mu _{0}\right) \frac{\partial \mu _{2}%
}{\partial r} \nonumber\\
&&+\frac{e{^{-2\,\mu _{1}}}}{12}\left( 10\frac{\partial }{\partial r}\left(
\mu _{0}-\mu _{1}\right) {\frac{\partial \mu _{0}}{\partial r}}+\frac{%
\partial }{\partial r}\left( \mu _{3}-2\,\mu _{2}-\mu _{1}+\,\mu _{0}\right)
{\frac{\partial \mu _{3}}{\partial r}}\right) +\frac{i\cos \theta }{2}{\frac{%
\partial \omega }{\partial r}}e{^{-\,\mu _{0}-\,\mu _{1}-\mu _{2}+\mu _{3}}}
\nonumber\\
&&+\frac{e{^{-2\,\mu _{1}}}}{12}\left( {\frac{\partial ^{2}}{\partial {r}^{2}%
}(}\mu _{3}+10\,\mu _{0}+\,\mu _{2})+\frac{2}{r}\,\frac{\partial }{\partial r%
}\left( {\,\mu _{0}-\,\mu _{1}}\right) -\frac{2}{r^{2}}\right) +\frac{{e{%
^{-2\mu _{2}}}}}{12{r}^{2}}\,{\frac{\partial }{\partial \theta }\left( \mu
_{1}+\mu _{3}-\mu _{2}+10\,\mu _{0}\right) {\frac{\partial \mu _{1}}{%
\partial \theta }}} \nonumber\\
&&{+\frac{{e{^{-2\mu _{2}}}}}{12{r}^{2}}\left( \frac{\partial }{\partial
\theta }\left( 2\,\mu _{2}-2\,\mu _{3}+\mu _{0}\right) {\frac{\partial \mu
_{3}}{\partial \theta }}+\frac{\partial }{\partial \theta }\left( \mu
_{0}-\mu _{2}\right) {\frac{\partial \mu _{0}}{\partial \theta }}\right) }+{%
\frac{{\cot \theta }}{12{r}^{2}}\left( \,{\frac{\partial }{\partial \theta }%
(-4\,\mu _{3}+2\,\mu _{2}}+{\mu }_{{0}}{+\mu _{1})}\right) }e{{^{-2\mu _{2}}}%
}\nonumber \\
&&{+\frac{{e{^{-2\mu _{2}}}}}{12{r}^{2}}}\left( {{\frac{\partial ^{2}}{%
\partial {\theta }^{2}}(}\mu _{0}-2\,\mu _{3}+\mu _{1})+2}\right) +\frac{%
i\sin \theta }{4}\left( \frac{\partial }{\partial r}\left( \mu _{3}-\mu
_{2}\right) {\frac{\partial \omega }{\partial \theta }}+\frac{\partial }{%
\partial \theta }\left( 2\mu _{3}-\mu _{0}-\mu _{1}\right) {\frac{\partial
\omega }{\partial r}}\right) e{^{-\,\mu _{0}-\,\mu _{1}-\mu _{2}+\mu _{3}}}\nonumber\\
&&+{\frac{i\sin \theta }{4}\frac{\partial ^{2}\omega }{\partial \theta
\partial r}}e{^{-\,\mu _{0}-\,\mu _{1}-\mu _{2}+\mu _{3}}},\\
\psi _{3} &=&-\frac{r\sin ^{2}\theta \,}{2}{\frac{\partial \omega }{\partial
\theta }\frac{\partial \omega }{\partial r}}e{^{-2\,\mu _{0}-\mu _{1}-\,\mu
_{2}+{2\,\mu _{3}}}}-\frac{2}{8{r}^{2}}\,{\frac{\partial }{\partial \theta }%
(\mu _{1}+3\,\mu _{0})e{^{-\,\mu _{1}-\,\mu _{2}}}-}\frac{{3i\cos \theta }}{%
8r}{{\frac{\partial {\omega }}{\partial \theta }}e{^{-\,\mu _{0}-2\,\mu
_{2}+\mu _{3}}}} \nonumber\\
&&+\frac{{e{^{-\,\mu _{1}-\,\mu _{2}}}}}{4r}{\frac{\partial ^{2}}{\partial
\theta \partial r}(}\mu _{3}+3\,\mu _{0})+\frac{i\sin \theta }{8r}{\left( {%
\frac{\partial }{\partial \theta }(}\mu _{0}-3\,\mu _{1}+{\mu _{2}-3\,\mu
_{3})}\frac{\partial \omega }{\partial \theta }-{\frac{\partial ^{2}\omega }{%
\partial {\theta }^{2}}}\right) e^{-\,\mu _{0}-2\,\mu _{2}+\mu _{3}}}\nonumber \\
&&+\frac{{e{^{-\,\mu _{1}-\,\mu _{2}}}}}{4r}{\,\left( -\frac{\partial }{%
\partial r}\left( \mu _{3}+3\,\mu _{0}\right) {\frac{\partial \mu _{1}}{%
\partial \theta }}+\frac{\partial }{\partial r}\left( \mu _{3}-\mu
_{2}\right) {\frac{\partial \mu _{3}}{\partial \theta }}+3\,\frac{\partial }{%
\partial r}\left( \mu _{0}-\mu _{2}\right) {\frac{\partial \mu _{0}}{%
\partial \theta }}\right) } \nonumber\\
&&+\frac{ir\sin \theta }{8}\left( \frac{\partial }{\partial r}\left( 3\mu
_{0}+3\mu _{1}-\mu _{2}-{9\,}\mu _{3}\right) \frac{\partial \omega }{%
\partial r}-3{\frac{\partial ^{2}\omega }{\partial {r}^{2}}}-\frac{10}{r}\,{%
\frac{\partial \omega }{\partial r}}\right) e{^{-2\mu _{0}-2\,\mu _{1}+\mu
_{3}}} \nonumber\\
&&+\frac{{\cot \left( \theta \right) }}{4r}\,{\frac{\partial }{\partial r}%
\left( \mu _{3}-\mu _{2}\right) e^{-2\mu _{2}-\mu _{0}-\mu _{1}}},\\
\psi _{4} &=&-\frac{\sin ^{2}\theta }{4}\,\left( {\frac{\partial \omega }{%
\partial \theta }}\right) ^{2}e{^{-2\,\mu _{0}-2\,\mu _{2}+2\,\mu _{3}}}+%
\frac{e{^{-2\,\mu _{1}}}}{4}\left( \frac{\partial }{\partial r}\left( \,\mu
_{3}-\,\mu _{1}-\,\mu _{0}\right) {\frac{\partial \mu _{3}}{\partial r}}+\,%
\frac{\partial }{\partial r}\left( \mu _{1}-\mu _{2}+\,\mu _{0}\right) {%
\frac{\partial \mu _{2}}{\partial r}}\right)  \nonumber\\
&&-\frac{e{^{-2\,\mu _{2}}}}{{r}^{2}}\,\left( \frac{\partial }{\partial
\theta }\left( {\mu _{1}-\mu _{3}-\mu _{2}}\right) {{\frac{\partial {\mu _{1}%
}}{\partial \theta }}+}\frac{\partial }{\partial \theta }\left( {\mu _{3}{%
-\mu _{0}}+\mu _{2}}\right) {{\frac{\partial \mu _{0}}{\partial \theta }}+{%
\frac{\partial ^{2}}{\partial {\theta }^{2}}(}\mu _{1}-\mu _{0})}\right)  \nonumber\\
&&+\frac{e{^{-2\,\mu _{1}}}}{4}\,\left( {\frac{\partial ^{2}}{\partial {r}%
^{2}}(}\mu _{3}-\,\mu _{2})-\frac{2}{r}\,{\,{\frac{\partial }{\partial r}(}%
\mu _{2}-\,\mu _{3})}\right) +\frac{{\cot \left( \theta \right) }}{4r^{2}}\,{%
\frac{\partial }{\partial \theta }\left( \mu _{1}-\mu _{0}\right) }e{%
^{-2\,\mu _{2}}} \nonumber\\
&&+\frac{i\sin \theta }{4}\left( \frac{\partial }{\partial r}\left( \mu
_{2}-3\mu _{3}+2\mu _{0}\right) {\frac{\partial \omega }{\partial \theta }+}%
\frac{\partial }{\partial \theta }\left( \mu _{1}-\mu _{0}\right) {\frac{%
\partial \omega }{\partial r}-\frac{\partial ^{2}\omega }{\partial \theta
\partial r}}-\frac{2}{r}{{\frac{\partial {\omega }}{\partial \theta }}}%
\right) e{^{-\,\mu _{0}-\,\mu _{1}-\,\mu _{2}+\mu _{3}}}.
\end{eqnarray}
\end{widetext}
Finally, we show the three Maxwell field scalars and the twelve spin coefficients defined in the NP approach,
\begin{widetext}
\begin{align}
\Phi _{0} &=-\frac{1}{2r}\,{\left( {\frac{\partial A_{\theta }}{\partial r}}%
-{\frac{\partial A_{r}}{\partial \theta }}\right) e{{^{-\mu 1-\mu 2}}}-\frac{%
1}{2r}\left( {\frac{\partial A_{\varphi }}{\partial \theta }}\omega -{\frac{%
\partial \chi }{\partial \theta }}\right) e{{^{-\mu 0-\mu 2}}}}
+\frac{i\csc \theta }{2r}{{\frac{\partial A_{\varphi }}{\partial r}}e{^{-\mu 1-\mu 3}}},
\end{align}
\begin{align}
\Phi _{1}&={\left( \omega {\frac{\partial A_{\varphi }}{\partial r}}-{\frac{%
\partial \chi }{\partial r}}\right) e{^{-\mu 0-\mu 1}}}+\frac{i\csc \theta }{%
2{r}^{2}}{{\frac{\partial A_{\varphi }}{\partial \theta }}e{^{-\mu 2-\mu 3}}},\\
\Phi _{2} &=\frac{1}{2r}\,{\left( {\frac{\partial A_{\varphi }}{\partial
\theta }}\omega -{\frac{\partial \chi }{\partial \theta }}\right) e{{^{-\mu
0-\mu 2}}}+\frac{1}{2r}\left( {\frac{\partial A_{r}}{\partial \theta }}-{%
\frac{\partial A_{\theta }}{\partial r}}\right) e{{^{-\mu 1-\mu 2}}}}
-\frac{i\csc \theta }{2r}{{\frac{\partial A_{\varphi }}{\partial r}e^{-\mu1-\mu 3}}},\\
\sigma &=\lambda =\frac{\sqrt{2}}{4}\,\left( i\sin \theta {\frac{\partial
\omega }{\partial \theta }}e{^{-\mu_0-\mu_2+\mu_3}}+{\frac{\partial }{%
\partial r}(}\mu_2-\mu_3)e{^{-\mu_1}}\right), \\
\rho &=\mu =\frac{\sqrt{2}}{4}\,{\left( {\frac{\partial }{\partial r}(}\mu
_{2}+\mu _{3})+\frac{2}{r}\right) e{^{-\mu _{1}}}},\\
\kappa &=-\nu =\frac{\sqrt{2}}{4r}\,\frac{\partial }{\partial \theta }{\left(
\mu _{0}-\mu _{1}\right) e{^{-\mu _{2}}}}, \\
\tau &= -\pi=-\frac{\sqrt{2}}{4r}{\left( i\sin \theta {{r}^{2}\frac{\partial
\omega }{\partial r}}e{{^{-\mu _{0}-\mu _{1}+\mu _{3}}}}-\frac{\partial }{%
\partial \theta }\left( \mu _{0}+\mu _{1}\right) e{{^{-\mu _{2}}}}\right) }, \\
\alpha &=-\beta =\frac{i\sqrt{2}r\sin \theta }{8}{\frac{\partial \omega }{%
\partial r}}e{^{-\mu _{0}-\mu _{1}+\mu _{3}}}+\frac{\sqrt{2}}{4r}e{{^{-\mu
_{2}}}}\left( {{\frac{\partial {\mu }_{3}}{\partial \theta }}}+{{\cot }%
\theta }\right), \\
\gamma &=\epsilon =-\frac{\sqrt{2}}{8}\left( i\sin \theta {\frac{\partial
\omega }{\partial \theta }}e{^{-\mu _{0}-\mu _{2}+\mu _{3}}}+2\,e{^{-\mu
_{1}}\frac{\partial \mu _{0}}{\partial r}}\right).
\end{align}
\end{widetext}


\end{document}